\documentclass[preprint2]{aastex}
\usepackage{natbib}
\usepackage{lscape}
\usepackage{wasysym}
\usepackage{txfonts}
\usepackage{graphicx}
\usepackage{hyperref}
\bibpunct{(}{)}{;}{a}{}{,}
\usepackage{longtable}

\usepackage{afterpage}

\def\ms{\hbox{\,m\,s$^{-1}$}}         
\def\m2s2{\hbox{\,m$^{2}$\,s$^{-2}$}} 
\def\kms{\hbox{\,km\,s$^{-1}$}}       

\newcommand{\titleast}{\ast}
\newcommand{\titlestar}{\star}

\shorttitle{Stellar inclination from stellar activity.}
\shortauthors{X. Dumusque}

\begin{document}

\title{
Deriving stellar inclination of slow rotators using stellar activity.\altaffilmark{\titleast}}

\author{X. Dumusque\altaffilmark{1}\altaffilmark{\titlestar}} 

\altaffiltext{1}{Harvard-Smithsonian Center for Astrophysics, 60 Garden Street, Cambridge, Massachusetts 02138, USA}

\altaffiltext{$\star$}
{Swiss National Science Foundation Fellow; xdumusque@cfa.harvard.edu}

\altaffiltext{$\ast$}
{Based on observations made with the \emph{MOST} satellite, the HARPS instrument on the ESO 3.6-m telescope at La Silla Observatory (Chile), and the SOPHIE instrument at the Observatoire de Haute Provence (France).}

\begin{abstract}
Stellar inclination is an important parameter for many astrophysical studies. Although different techniques allow us to estimate stellar inclinationt for fast rotators, it becomes much more difficult when stars are rotating slower than $\sim2$-2.5 \kms.
By using the new activity simulation SOAP 2.0 that can reproduce the photometric and spectroscopic variations induced by stellar activity, we are able to fit observations of solar-type stars and derive their inclination.
For HD189733, we estimate the stellar inclination to be $i=84^{+6}_{-20}$ degrees, which implies a star-planet obliquity of $\psi=4^{+18}_{-4}$ considering previous measurements of the spin-orbit angle.
For $\alpha$ Cen B, we derive an inclination of $i=45^{+9}_{-19}$, which implies that the rotational spin of the star is not aligned with the orbital spin of the $\alpha$ Cen binary system.
In addition, assuming that $\alpha$ Cen Bb is aligned with its host star, no transit would occur. 
The inclination of $\alpha$ Cen B can be measured using 40 radial-velocity measurements, which is remarkable given that the projected rotational velocity of the star is smaller than 1.15\kms.
\end{abstract}


\section{Introduction} \label{sect:1}

In many different fields of astrophysics, obtaining stellar inclination is often a critical step for further modeling. Reconstructing spots maps from Doppler imaging requires stellar inclination to recover the latitude of spots on the stellar surface \citep[e.g.][]{Vogt-1987}. This inclination can however be constrained during the Doppler imaging fitting process itself if the signal-to-noise ratio of the data is sufficient because a wrong inclination produces systematic errors in the spot map \citep[][]{Rice-2000}. 
When studying the large scale topology of magnetic fields using Zeeman Doppler Imaging measurements, the stellar inclination have also to be chosen to lift the degeneracy between magnetic field configuration and inclination of the star \citep[e.g.][]{Donati-2006}. Last but not least, stellar inclination is required to get the true obliquity of a transiting planet orbiting its host star, i.e the angle between the orbital angular momentum and the stellar rotation axis \citep[e.g.][]{Fabrycky-2009}. The obliquity can be obtained from the stellar inclination if a measurement of the sky projected obliquity angle, most commonly called spin-orbit angle, can be performed \citep[e.g.][]{Benomar-2014,Lund-2014}. This sky projected obliquity angle is generally measured using the Rossiter-McLaughlin effect \citep[][]{Queloz-2000,Rossiter-1924,McLaughlin-1924}.
Obtaining the obliquity is of prime importance to distinguish between the various migration scenarios that have been proposed to explain the existence of close-in giant-like hot Jupiters or hot Neptunes. Indeed, theories like disk-migration predict a rather small misalignment between stellar spin and planetary orbital axes \citep[e.g.][]{Lin-1996}, while theories like planet-planet scattering or migration produced by Kozai cycles in combination with tidal circularization predict a very wide range of obliquity angles \citep[e.g.][]{Nagasawa-2011,Fabrycky-2007,Wu-2003}. Note however that stellar magnetic fields can induce an obliquity between the stellar spin and the planetary orbital angular momentum even in the disk migration scenario \citep[see][]{Lai-2011}.

\section{Deriving stellar inclination and obliquity} \label{sect:2}

\subsection{Measuring the projected rotational velocity with spectroscopy} \label{sect:2-1}

The most common technique to derive stellar inclination is first to estimate the stellar projected rotational velocity $v\sin{i}$, where $v$ is the equatorial rotational velocity of the star, and $i$ its inclination. This can be done by matching an observed stellar spectrum to a grid of synthetic model spectra \citep[e.g. SPC and SME codes,][]{Buchhave-2012,Valenti-2005,Valenti-1996}. Then an estimation of the rotational period of the star $P_\mathrm{rot}$, using the photometric light curve \citep[][]{Hirano-2014,Hirano-2012}, rotation-age-mass correlations \citep[][]{Schlaufman-2010} or the chromosheric calcium index modulation \citep[e.g.][]{Lendl-2014}, gives us $v$ and thus the stellar inclination\footnote{using the formula $v=2\pi R_{\star} / P_{\mathrm{rot}}$, where $R_{\star}$ is the stellar radius generally derived from evolutionary tracks or asteroseismology.}.

Another possibility to get $v\sin{i}$ is to study the width of the Cross Correlated Function (CCF) obtained after cross-correlating a stellar spectrum with a synthetic template \citep[][]{Pepe-2002,Baranne-1996}. The CCF can be considered as an average line of the target spectrum, and therefore carries information on the stellar atmospheric parameters (like the global abundances, thermal broadening, pressure broadening or micro-turbulence), the macroturbulence, the projected rotational velocity \citep[][]{Gray-2008}, and the instrumental profile. 

The Gaussian width of a weak spectral line of a "non-rotating" star depends mostly on its spectral type and luminosity class. Therefore by observing with the same instrument, i.e. same instrumental profile, dwarfs that have a similar \emph{B-V} color, the only parameter affecting the width of the CCF is the projected rotational velocity of the stars. Stars that have the minimum width will be associated to "non-rotators", i.e. stars that are seen pole-on, and all excess width will be associated to non-zero projected rotational velocity \citep[][]{Boisse-2010,Santos-2002b,Queloz-1998,Benz-1981}.

Deriving stellar inclination from a $v\sin{i}$ measurement obtained either by matching stellar spectra or using the width of the CCF is limited by the instrumental resolution of the spectrograph used. Even for high-resolution instruments like HARPS, HARPS-N, HIRES, it is difficult to measure precise $v\sin{i}$ for stars rotating slower than 2 - 2.5 \kms.

If the star is hosting a transiting planet, the $v\sin{i}$ can be obtained by measuring the Rossiter-McLaughlin effect. By masking part of the stellar disc in rotation, the transiting planet creates an anomaly in the radial velocity (RV) curve that is proportional to $v\sin{i}$ \citep[][]{Winn-2010,Triaud-2009,Queloz-2000}. However, the use of different models estimating the Rossiter-McLaughlin effect can lead to different results for the shape of the RV anomaly and therefore different $v\sin{i}$ determinations \citep[][]{Boue-2013}. In the special case when the spin-orbit angle measured using the Rossiter-McLaughlin effect is close to zero, there is a high probability that the star is seen equator-on, and thus the obliquity of the planet is close to zero. Indeed, if the spin-orbit angle is close to zero, the stellar inclination can be different from 90 degrees only in the plane perpendicular to the planetary orbit that contains the line of sight. The probability of the stellar spin being in this plane is very small compared to all the possible orientations and therefore there is a high probability that the stellar inclination is close to 90 degrees.

\subsection{Measuring rotational splitting using asteroseismology}  \label{sect:2-2}

Another possible way to measure stellar inclination is by observing rotational splitting of the oscillation modes of the star. Detailed descriptions of the principles of this method based on asteroseismology may be found in \citep[][]{Ballot-2008,Ballot-2006,Gizon-2003b}.

In the absence of rotation, the frequency of a mode depends only on its radial spherical harmonic order $n$ and its degree $l$. Modes are $(2l + 1)-$times degenerate among the azimuthal spherical harmonic order $m$. This degeneracy is removed by breaking the spherical symmetry, especially by rotation. For geometrical reasons, only modes with a $l\le3$ have a sufficient amplitude to be visible in an oscillation spectrum due to the integration of the signal over the entire stellar disc. For $l=1$, each multiplet (n,l) will have three peaks in the power spectrum and the relative heights between them allows us to constrain the stellar inclination \citep[][]{Huber-2013}.

An asteroseismic analysis requires bright targets and long-duration, high-cadence space-based photometric time series to give the requisite signal-to-noise and frequency resolution for extracting clear signatures of rotation from the oscillation spectrum, and hence the stellar inclination angle. Up to know only a few inclination studies with asteroseismology were carried out on solar-type stars observed with \emph{Kepler} as signals are faint for these type of targets \citep[][]{Van-Eylen-2014,Chaplin-2013,Huber-2013}.

\subsection{Spot occultation during transit}  \label{sect:2-3}

The passage of a transiting planet in front of a star can map spots on its surface, which can be used to infer the stellar inclination depending if the planet mask the same spot in consecutive transits or not \citep[e.g.][]{Sanchis-Ojeda-2012,Desert-2011,Sanchis-Ojeda-2011,Nutzman-2011}.

This technique requires space-based photometry in high-cadence mode during several consecutive transits, active stars presenting big spots, and relatively big planets to map sufficiently the stellar surface. In addition, the period of the transiting planet must be much shorter than the rotational period of the star and than the spot lifetime so that the occultation signal stays in phase for a few consecutive transits.

\subsection{Other methods to derive stellar inclination}

For extremely fast rotators, the obliquity of the system can be obtain using the gravity darkening signature. This has been done so far for the KOI-13 and KOI-368 systems \citep[][]{Ahlers-2014,Szabo-2011,Barnes-2011,Barnes-2009}.

Stellar inclination can also be obtain in special cases using the beaming effect \citep[][]{Shporer-2012,Groot-2012}.

\section{Fitting stellar activity to derive stellar inclination}

In this section, we present a new technique to derive stellar inclination using the photometric and spectroscopic variation induced by short-term activity, i.e. modulations induced by the presence of active regions on the stellar surface. In principle, the inclination of the stellar spin axis can be extracted from photometric and spectroscopic measurements when one major active region, spot or plage, is dominating the activity signal. With the photometry alone, it is possible to study the amplitude of the light modulation and the duration of the flux anomaly produced by an active region that is in view only for a fraction of the rotational phase. This provides information on a combination of the stellar spin inclination, the active region latitude and its area\footnote{The active region size is degenerated with the active region contrast if the active region temperature is not fixed (see Section \ref{sect:6}).}. These three parameters can not be characterized individually and a third observable is required to lift the degeneracy between them. This third observable can come from spectroscopy measurements, for which the amplitude of the RV signal can be used.
Estimating the stellar inclination that way requires to know the equatorial rotational velocity of the star, which is often derived from the stellar radius and the rotational modulation seen in photometry. In conclusion, if the the temperature of the active region is fixed and the equatorial velocity is known, we can extract information on the stellar spin axis and the active region latitude and area.

Several precedent attempts have tried to derive the stellar inclination by using the photometric and RV information \citep[][]{Boisse-2012b,Lanza-2011b}, however a model that estimates in a proper way the photometric and spectroscopic variations of active region is required.
In this paper, we use the results of the SOAP 2.0 code recently published \citep[][]{Dumusque-2014b} that estimates the effect of spots and plages based on spectroscopic observations of the Sun. This code allows us to reproduce the activity-induced variation seen in photometry, RV, bisector span (BIS SPAN) and Full Width at Half Maximum (FWHM) of the CCF. We show in the following two examples that fitting simultaneously all these observables lift the degeneracy between stellar inclination, active region size and latitude, and in the end the stellar inclination can be inferred.

\subsection{HD189733}\label{sect:5-1}

To illustrate how we can derive the stellar inclination of a star from fitting its stellar activity variations, we will first use a rather active star that rotates moderately fast, for which spots should be the dominant active regions \citep[][]{Shapiro-2014,Lockwood-2007}. In that case, the flux effect should explain the major part of the photometric, RV, BIS SPAN and FWHM activity-induced variations \citep[][]{Dumusque-2014b}. As done in the original SOAP paper \citep[][]{Boisse-2012b}, we use HD189733 as a benchmark. 
This star host a hot Jupiter \citep[][]{Bouchy-2005c} and is an active star variable at the percent level. Activity have been detected photometrically \citep[][]{Winn-2007,Croll-2007}, 
but also in X-ray \citep[][]{Poppenhaeger-2013a} and in calcium activity index \citep[][]{Boisse-2009,Moutou-2007}. Studying the modulation of the photometric activity signal, \citet{Henry-2008} find a rotational period for the star of 11.95 days.
\begin{deluxetable}{ccccc}
	\tabletypesize{\scriptsize}
	\tablewidth{14cm} 
	\tablecaption{Parameters of HD189733 and $\alpha$ Cen B. \label{tab:4}}
		\startdata 
			\tableline
			&  \multicolumn{2}{c}{HD189733} & \multicolumn{2}{c}{$\alpha$ Cen B}\\
			Parameter & Value & Ref. & Value & Ref. \\
			\tableline
			Radius ($R_{\odot}$) & $0.766\pm 0.01$ & \citet{Triaud-2009} & $0.863\pm 0.005$ & \citet{Kervella-2003}\\
			\tableline
			T$_{\mathrm{eff}} (K)$ & $5040\pm 50$ & \citet{Torres-2008} & $5214\pm 33$ & \citet{Dumusque-2012}\\
			$[$Fe$/$H$]$ & $-0.03\pm 0.08$ & \citet{Torres-2008} & $0.19\pm 0.09$ & \citet{Santos-2005a}\\
			logg & $4.59\pm 0.02$ &  \citet{Torres-2008} & $4.37\pm 0.12$ & \citet{Santos-2005a}\\
			Limb darkening $\gamma_1$ & 0.7787 &  \citet{Claret-2004}& 0.7207 & \citet{Claret-2004}\\
			Limb darkening $\gamma_2$ & 0.0549 & \citet{Claret-2004}& 0.1054 & \citet{Claret-2004}\\
			\tableline
			Active region type & Spot& \citet{Dumusque-2014b} & Plage & \citet{Dumusque-2014b}\\
			$\Delta$T active region (K) & $-663$ & \citet{Meunier-2010a} & $250.9-407.7\cos{\theta}+190.9\cos^2\theta$ & \citet{Meunier-2010a}\\
			\tableline
			Instrument Resolution & 75000 & SOPHIE (High Res. mode) & 115000 & HARPS\\
		\enddata
	\tablecomments{Parameters of HD189733 and $\alpha$ Cen B used when fitting the observed data. The effective temperature, metallicity and gravity are only used to derive the limb darkening parameters. $\theta$ is the angle between the normal to the stellar surface and the observer ($\theta=0$ at the stellar disc center and $\pi/2$ at the limbs).}
\end{deluxetable}

\begin{figure*}
\begin{center}
\includegraphics[width=16cm]{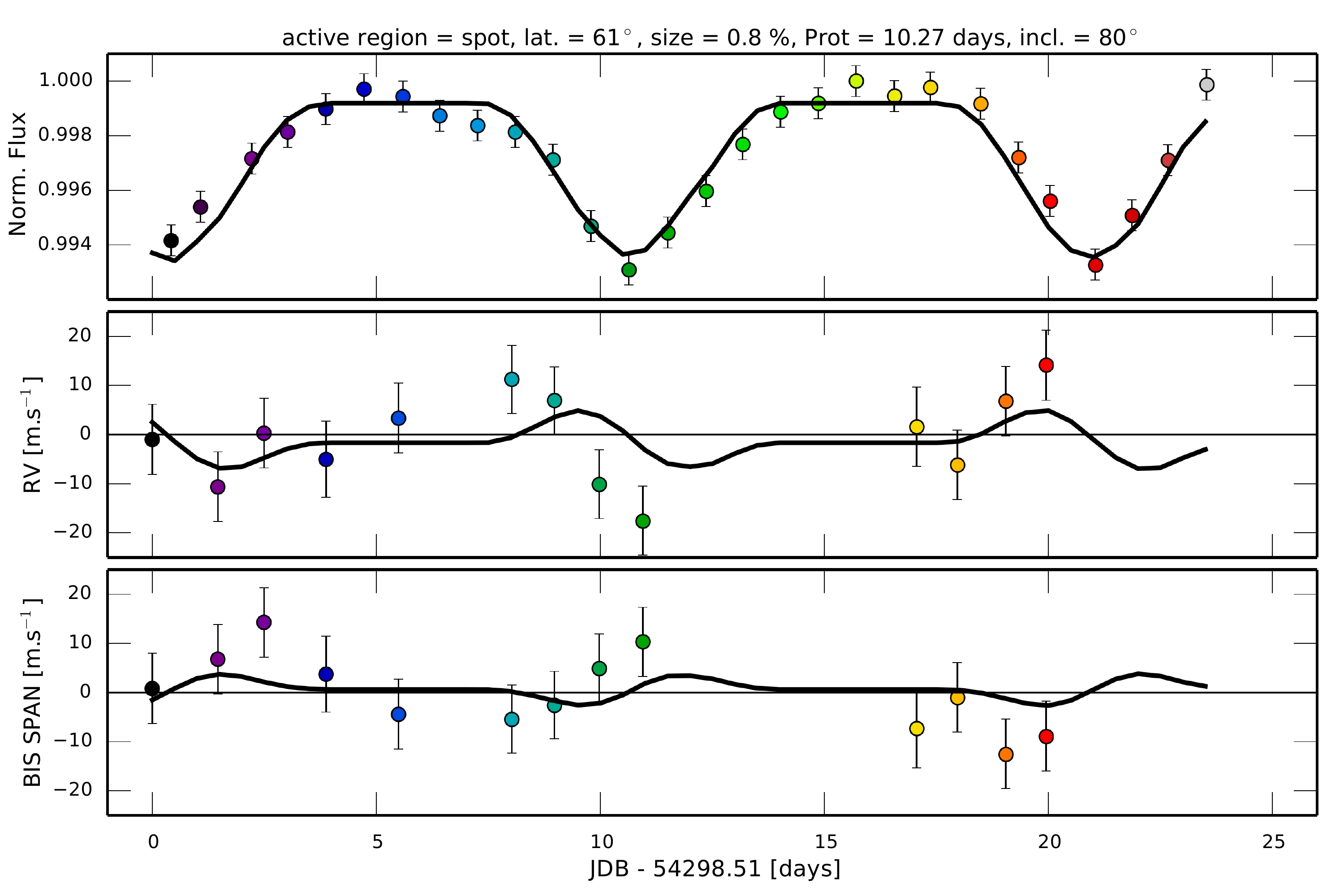}
\caption{Photometric, RV, and BIS SPAN variations and best fit (black continuous line) of the activity-induced signal observed on HD189733. Our best fit to the data corresponds to a 0.8\% spot that can be found at a latitude of 61 degrees. The star rotates in 10.27 days and is seen nearly equator-on with an inclination of 80 degrees.}
\label{fig:3-0}
\end{center}
\end{figure*}

HD189733 has been observed in July 2007 simultaneously in spectroscopy with SOPHIE at the Observatoire de Haute Provence in France and in photometry with the \emph{MOST} satellite. 
We use these data, published in \citet{Aigrain-2012}, \citet{Lanza-2011b} and \citet{Boisse-2009}, and the SOAP 2.0 code to fit with a Markov Chain Monte Carlo (MCMC) algorithm \citep[based on PyMC,][]{Patil-2010} the observed variations in photometry, RV and BIS SPAN. The FWHM of HD189733 for the same period exhibits a peak-to-peak amplitude of 135\ms, which is unlikely to be due only to activity-induced variations and could be induced by moon contamination (see following discussion). We therefore decided not to include the FWHM in our fitting procedure. Note that only one active region will be fitted to the photometric and spectroscopic data, which is justified given the regular photometric variation over two rotational periods\footnote{A linear trend was fitted to the \emph{MOST} photometric data to account for an instrumental drift or a long-term activity variation not related to rotational modulation.} (see Figure \ref{fig:3-0}). The original data of \emph{MOST} and SOPHIE are binned over one day to average out shorter stellar signals like oscillations and granulation, and to reduce the total number of points, which reduces the computational time of each iteration of the MCMC.
To fit the activity-induced variation, the planetary signal of HD189733b has been removed from the RVs using the solution of \citet{Boisse-2009}, and the residual RVs and the BIS SPAN have been centered on zero using a weighted mean.

The following input parameters have to be given before running the MCMC: the radius and the limb darkening coefficients for the star, the resolution of the spectrograph used to obtain the RVs, the active region type, and the active region temperature. The values used for these parameters are given in Table \ref{tab:4}. 
To select the type of active region, we compared the photometric and RV amplitudes of the activity signal. If a plage was at the origin of the photometric variation, the RV and BIS SPAN variations would be much larger, and thus we decided to use a spot to fit the data \citep[][]{Dumusque-2014b}.
\citet{Pont-2013} also found the presence of spots on HD189733 and tried to estimate the temperature difference compared to the photosphere of the spots occulted during the transit of HD189733b. They arrived to the conclusion that this temperature difference is $-750\pm250\,K$, which is compatible with the $-663\,K$ \citep[][]{Meunier-2010a} used in SOAP 2.0. We therefore decided to use this later value.

The free parameters fitted with the MCMC are the active region longitude, latitude $\phi$ and size $S$, the stellar rotational period and inclination $i$, in addition to a stellar jitter term for the photometry and another one for the spectroscopy. The size $S$ is defined as the fraction of the surface of the visible hemisphere covered by the active region. Because $S$, $\phi$ and $i$ are correlated due to geometrical symmetries and projections, the following empirical change of variables: 
\begin{eqnarray} \label{eq:5-0}
\alpha &=& \sqrt{S}\,sin(i)\,sin(\phi) \qquad S = \alpha^2+\beta^2+\gamma^2, \nonumber \\
\beta &=&  \sqrt{S}\,sin(i)\,cos(\phi) \qquad i = cos^{-1}\left(\frac{\gamma}{\sqrt{S}}\right), \nonumber \\
\gamma &=& \sqrt{S}\,cos(i) \,\,\,\qquad \qquad \phi = tan^{-1}\left(\frac{\alpha}{\beta}\right),
\end{eqnarray}
is performed to reduce the correlation between these parameters, and therefore improve the efficiency of the MCMC. The photometric and spectroscopic jitter terms are quadratically added to the flux and the RV and BIS SPAN error bars\footnote{The error bars on the BIS SPAN is considered here the same as the ones on the RVs.} when maximizing the log likelihood, respectively.

Seven MCMC chains of 2$\times 10^6$ steps each is obtained to fit the observed data of HD189733 starting with random initial values within the following uniform priors:
\begin{eqnarray}
\alpha &=& [-\sqrt{S_{\mathrm{max}}},\sqrt{S_{\mathrm{max}}}] \\
\beta &=& [0,\sqrt{S_{\mathrm{max}}}] \nonumber\\
\gamma &=& [0,\sqrt{S_{\mathrm{max}}}] \nonumber\\
\mathrm{P}_{\mathrm{rot}} &=& [9,14]\nonumber\\
\mathrm{Longitude} &=& [-50,100]\nonumber\\
\mathrm{Jitter\,Flux} &=& [0,50\times med(\sigma_{Flux})]\nonumber\\
\mathrm{Jitter\,RV} &=& [0,10\times med(\sigma_{RV})],
\end{eqnarray}
where $S_{\mathrm{max}}$ is the maximum size allowed for the active region, fixed here at 50\%, and $med(\sigma_{Flux})$ and $med(\sigma_{RV})$ are the median of the flux and the RV error bars, respectively. To prevent symmetries, the inclination of the star is allowed to vary from 0 to 90 degrees and the latitude can take any value between $-90$ and $90$ degrees, which implies that $\beta > 0$ and $\gamma > 0$. Because $\alpha$, $\beta$ and $\gamma$ are still correlated between each other, despite the change of variable (see Eq. \ref{eq:5-0}), an adaptive Metropolis-Hasting step method is used to explore better the full parameter space \citep[][]{Haario-2001}. 

\begin{figure*}
\begin{center}
\includegraphics[width=16cm]{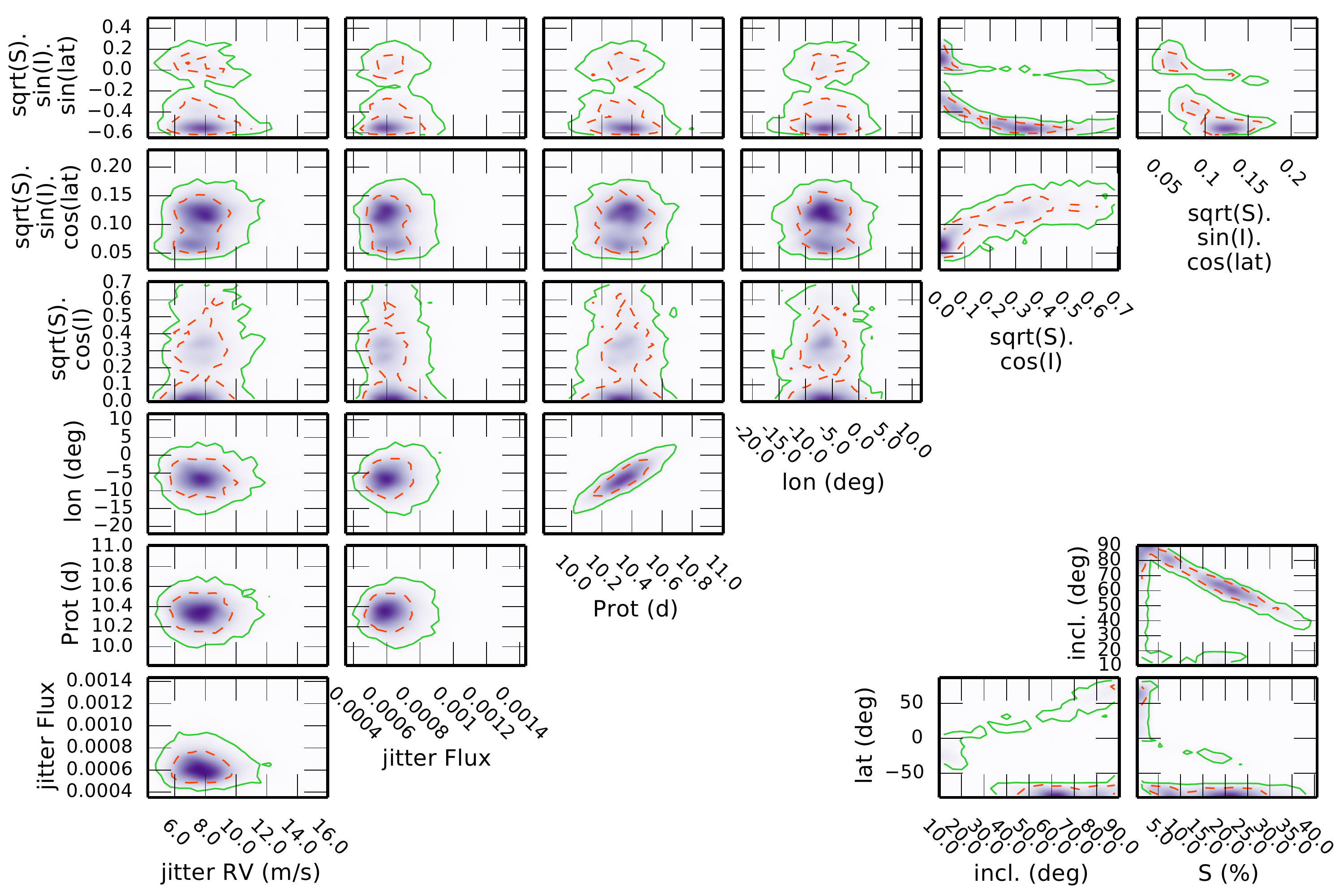}
\caption{Correlation between the different posterior distributions obtained from the MCMC fit to the data of HD189733 using an unconstrained prior for $\alpha$ and $\gamma$, which means that all spot latitudes and sizes are allowed. The $1-\sigma$ and $2-\sigma$ contours are shown in dashed red and continuous green lines, respectively.}
\label{fig:3-01}
\end{center}
\end{figure*}

A Gelman-Rubin test \citep[][]{Gelman-2004} on the seven chains gives a potential scale reduction better than 1.0049 for all parameters, proving the convergence and the proper mixing of the chains. With the first 2$\times 10^5$ steps rejected to remove the burn-in-period, and without any constrains on $\alpha$, the correlation between the posterior distributions of the fitted parameters in Figure \ref{fig:3-01} show that the data are compatible with a stellar inclination angle above 50 degrees, for which either we can have a big spot near the southern pole, or a smaller one in the northern hemisphere. These two solution are equivalent because it is not possible to differentiate between a northern and a southern spot at the same latitude when the star is close to equator on. The solution with the spot on the southern hemisphere is more likely, because in this case, the spot can grow to very large sizes as part of it will be out of view to the observer and therefore will not contribute to the signal. This can be seen when studying the inclination-size correlation plot. Note also that when the spot grows to very large sizes, the RV jitter goes to non-realistic values higher the 10\ms. To lift this degeneracy between a southern or a northern spot, and to prevent the spot from growing on the invisible part of the star, we selected the northern solution by imposing the following priors on $\alpha$ and $\gamma$:
\begin{eqnarray}
\alpha &=& [-0.1,\sqrt{S_{\mathrm{max}}}]\nonumber\\
\gamma &=& [0,0.2],
\end{eqnarray}
where $\alpha = -0.1$ is the delimitation between the southern and northern solutions (see the top row correlation plots for $\alpha$ in Figure \ref{fig:3-01}), and $\gamma = 0.2$ constrain the spot size to be smaller than 10\% for stellar inclination higher than 50 degrees.

\begin{figure*}
\begin{center}
\includegraphics[width=16cm]{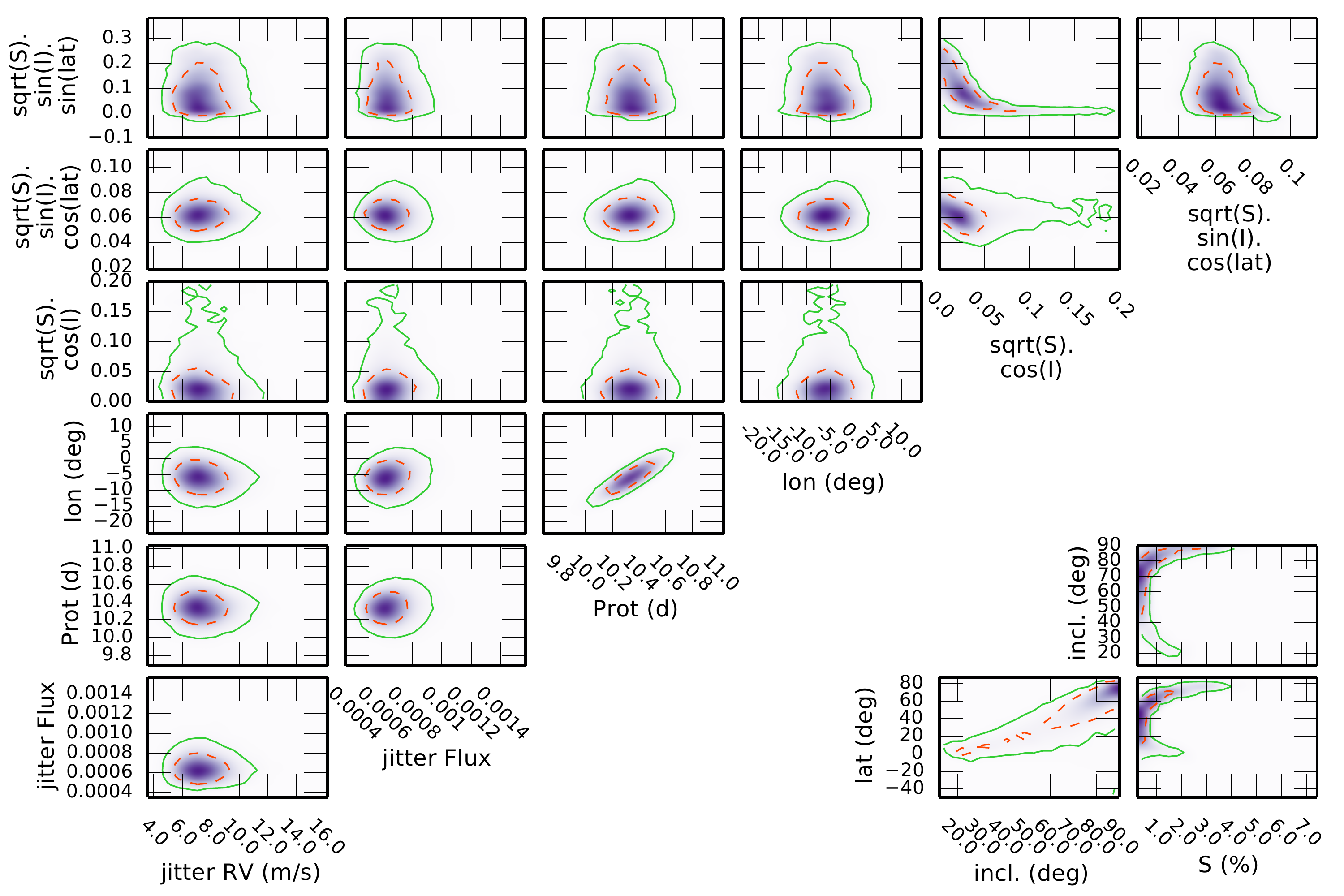}
\caption{Correlation between the different posterior distributions obtained from the MCMC fit to the data of HD189733 constraining $\alpha$ and $\gamma$ so that the spot is on the northern hemisphere and cannot grow in size on the hidden part of the star. The $1-\sigma$ and $2-\sigma$ contours are shown in dashed red and continuous green lines, respectively.}
\label{fig:3-1}
\end{center}
\end{figure*}

\begin{figure*}
\begin{center}
\includegraphics[width=16cm]{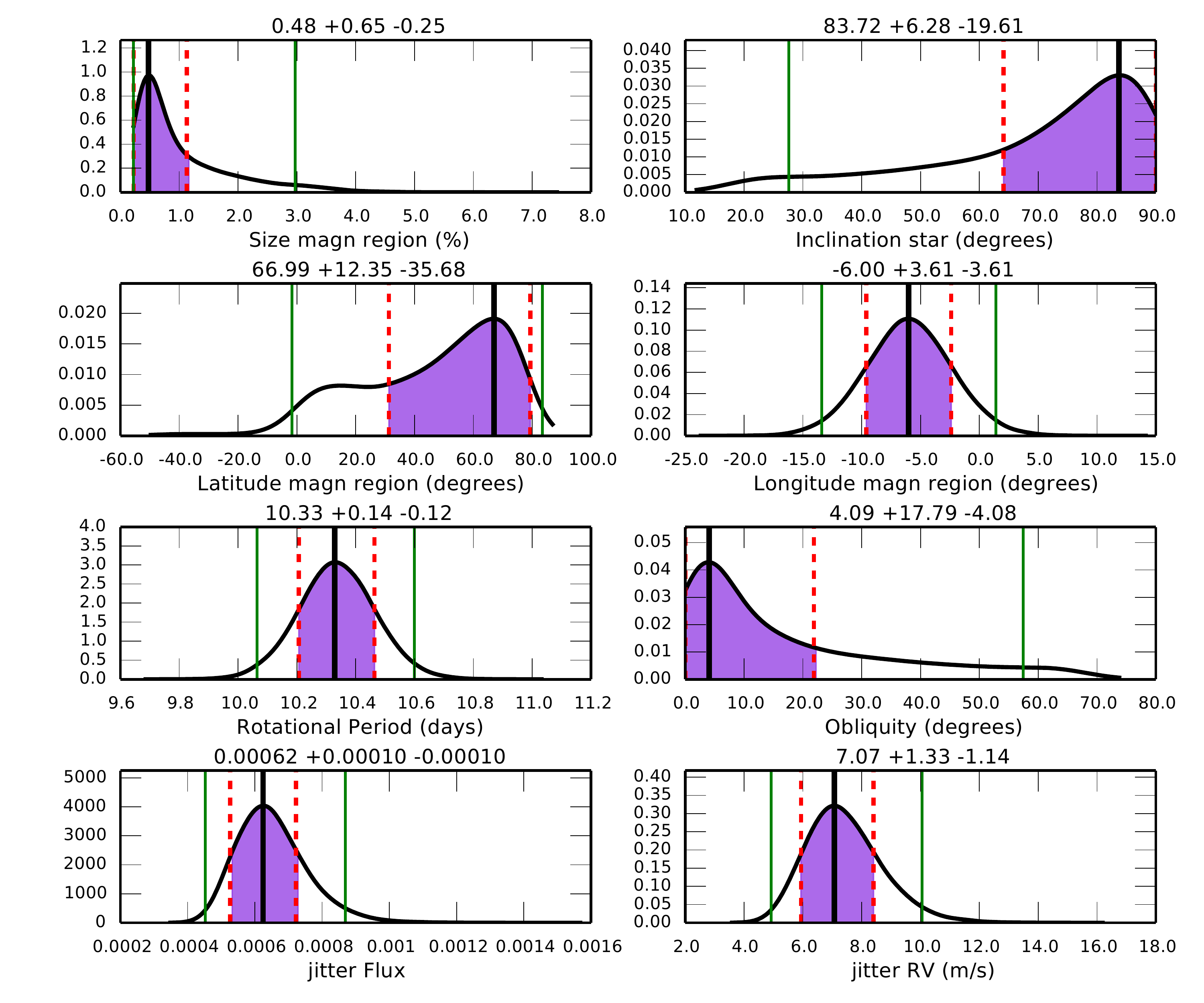}
\caption{Marginalized posterior distributions returned by our MCMC fit to the data of HD189733. The mode (black thick line) of the distribution for each parameter, with its $1-\sigma$ uncertainty (purple shaded regions delimited by red dashed lines) and $2-\sigma$ uncertainty (delimited by green thin lines), can be found on each plot. The title of each plot gives the value for the mode of the distribution and its $1-\sigma$ uncertainty.}
\label{fig:3-2}
\end{center}
\end{figure*}

We run a new MCMC chain with 2$\times 10^6$ steps with this new constrain on $\alpha$ and $\gamma$ and removed the first 2$\times 10^5$ to reject the burn-in-period.
Figure \ref{fig:3-1} shows the posterior correlations between the different fitted parameters and Figure \ref{fig:3-2} shows the marginalized posterior distributions of the parameters including the obliquity of the star-planet system. 
The best fitted solution maximizing the log likelihood is represented by the black curve in Figure \ref{fig:3-0}, and corresponds to a $\chi^2$ of 1.18 compared to 20.31 for a flat model.


The fit does not match the two RV measurements near BJD = 2454308.5 (10 days in the abscissa of Figure \ref{fig:3-0}).
In the studies by \citet{Aigrain-2012} and \citet{Lanza-2011b}, the same anomaly was reported using a spot model taking into account the flux effect and in some way the convective blueshift effect. 
 These two RV measurements were obtained near the full moon (BJD = 2454311.5), which can contaminate some spectra in case of clouds. \citet{Boisse-2009} removed strongly contaminated spectra from the observations, however, without simultaneous observation of the sky\footnote{the second fiber was illuminated by a thorium lamp for cross calibration, and not by the nearby sky.}, it is possible that some of the remaining spectra are slightly contaminated. Note that this contamination could be at the origin of the large peak-to-peak amplitude observed in the FWHM. In addition, as already discussed by \citet{Lanza-2011b}, flares could also be the cause of this anomaly, because the calcium activity index of HD189733 can sometimes vary on a very short timescale \citep[][]{Fares-2010,Moutou-2007}.

Removing the two bad points of the anomaly and considering only the spectroscopic data (RV and BIS SPAN), the reduced $\chi^2$ of the fit is 1.17 compared to 1.26 for a flat model, and the standard deviation of the RV residuals is 5.57\ms compared to 7.53\ms, which is an improvement of 5.06\ms. Although the improvement in $\chi^2$ only considering the spectroscopy is not very significant comparing our best fit model to a flat model, we have to note that photometry and spectroscopy are both fitted together and that photometry his much more constraining the fit than spectroscopy in this case. With this slight improvement in $\chi^2$ and the improvement in standard deviation, we are confident that our best fit reproduces better the data than a flat model.

The marginalized posteriors for the rotational period converges to $10.33^{+0.14}_{-0.12}$ days, which is smaller than the previous estimate of 11.95 days. The rotational period estimated by active regions gives the stellar rotational period at the latitude of these regions. Surface differential rotation can occur on solar-like stars which implies that different rotational period values can be obtained depending on the latitude of the active region responsible of the observed variation \citep[][]{Reinhold-2013b,Ammler-von-Eiff-2012,Reiners-2006,Barnes-2005}. Note that one of the first estimate for the rotational period of HD189733 was 13.4 days \citep[][]{Winn-2007}. 

The long tail in the stellar inclination-latitude correlation (see Figure \ref{fig:3-1}) reflects that stellar inclination, spot size and latitude are degenerated. 
However, fitting simultaneously the photometric, RV and BIS SPAN variations lift partially this degeneracy and the fit converges to a unique solution with an inclination of $i=84^{+6}_{-20}$ degrees, a spot latitude of $67^{+12}_{-36}$ degrees and a spot size of $0.5^{+0.7}_{-0.3}$\%. 
This solution for the stellar inclination implies that the star is seen nearly equator on, which is expected because the spin-orbit alignment measured using the Rossiter-McLaughlin effect is close to zero \citep[see Section \ref{sect:2-1}][]{Collier-Cameron-2010,Triaud-2009,Winn-2006}. 
Taking into account the impact parameter of the planet during the transit, \citet{Triaud-2009} derive an inclination of the orbital plane of the planet relative to the projection of the sky of $i_p=85.5\pm0.1$ degrees. Assuming a perfect alignment between the stellar spin and the orbital spin of the planet leads to a stellar inclination of $i=85.5\pm0.1$ degrees, in good agreement with our value found by fitting stellar activity. Therefore, it seems that fitting stellar activity using the results of the SOAP 2.0 code give us access to the stellar inclination.

According to \citet{Fabrycky-2009}, the obliquity is defined by:
\begin{eqnarray}
\cos{\psi} = \sin{i}cos{\lambda}\sin{i_p} + \cos{i}\cos{i_p},
\end{eqnarray}
where $\lambda$ is the spin-orbit angle, which can be measured by the Rossiter McLaughlin effect \citep[][]{Rossiter-1924,McLaughlin-1924}. With our stellar inclination posterior for HD189733, the spin-orbit angle and orbital plane of the planet found by \citet{Triaud-2009}, $\lambda=-0.85\pm0.32$ degrees and $i_p=85.5\pm0.1$ degrees, we estimate the obliquity of the system to be $\psi=4^{+18}_{-4}$ degrees (see Figure \ref{fig:3-2}). The alignement of HD189733b with its host star is expected given the effective temperature of HD189733 \citep[][]{Winn-2010b}.

One year before the \emph{MOST} and SOPHIE simultaneous observations, the Hubble Space Telescope (HST) observed three transits of HD189733b (in May and July 2006), and the data revealed that the planet was masking stellar spots during its passage in front of the stellar disc \citep[][]{Pont-2007}. Given the impact parameter of the planet $b=0.671\,R_{\star}$ and the planet to star radius ratio of 0.16 inferred at the time, plus the stellar inclination found here, the latitude of the spots observed by HST are estimated to be at $36^{+14}_{-11}$ degrees in latitude. 
This value is compatible with the latitude of the spot fitted in this paper, although we have large uncertainties. Note that the planet is also occulting spots in another study using a different data set \citep[][]{Pont-2013}.

The HST and \emph{MOST} data have been taken with a separation of one year, therefore the spot seen on the \emph{MOST} data is not necessarily one of the spots observed with HST, as spots evolve and disappear with time. On the Sun, spots appear at a preferred latitude, that varies with the magnetic cycle phase. During an eleven-year magnetic cycle, spots drift from $\sim40$ degrees in latitude to the equator. Assuming that the activity of HD189733 can be compared to the Sun, which seems reasonable given the results of \citet{Reiners-2006}, we expect only a small change in the preferred latitude of spots during a one year timescale. It is therefore not surprising that a spot latitude compatible with the HST observations is found.

\subsection{$\alpha$ Cen B}\label{sect:5-2}

The next step is to derive the stellar inclination for a slow rotator, for which the effect of plages is dominating the activity-induced variation \citep[][]{Shapiro-2014,Lockwood-2007}. Slow rotators are the main targets of RV surveys searching for small-mass planets because they tend to be less active, and therefore to exhibit a smaller activity-induced RV variation than rapid rotators. Many slow rotators have been observed using HARPS, HARPS-N and HIRES with the sufficient RV precision and cadence to study activity. However, to reduce at maximum the impact of activity when searching for planets, the observations are generally taken when these stars are at the minimum of their magnetic cycle, when only small active regions are present on the stellar surface. In this case, the activity-induced RV signal is at the level of the instrumental noise, and these data cannot be used to study the effect of activity on slow rotators. Nevertheless, a few data sets exist, and one of the best is the RV measurements that have been used to detect the closest planet to our Solar System orbiting $\alpha$ Cen B \citep[][]{Dumusque-2012}. This star has been observed between 2008 and 2011, during which the stellar activity level changed from minimum to maximum due to a solar-like magnetic cycle. In 2010, the data exhibit an important and extremely regular activity index variation \citep[in Ca II H and K,][]{Dumusque-2012} that can be modeled by a single major active region present on the stellar surface. To fit the activity-induced variation, the binary contribution of $\alpha$ Cen A has been removed from the raw RVs published in \citet{Dumusque-2012}, and the residual RVs, the BIS SPAN and the FWHM have been centered on zero using a weighted mean.

Looking at the data, we can see that the FWHM peak-to-peak amplitude is nearly four times larger than the RV peak-to-peak amplitude. Using the results of Section 4 in \citet{Dumusque-2014b}, this ratio between the amplitudes of the RV and the FWHM of the observed signal can be explained if a plage is dominating the activity-induced variations. The ratio between the RV, BIS SPAN and FWHM peak-to-peak amplitudes for $\alpha$ Cen B implies a $v\sin{i}\sim$1\,km\,s$^{-1}$ according to the SOAP 2.0 results for a plage \citep[see Figure 7 in][]{Dumusque-2014b}. With a rotational period of 37.8 days for $\alpha$ Cen B \citep[][]{Dumusque-2012} and a radius of 0.863 $R_{\odot}$ \citep[][]{Kervella-2003}, the stellar project rotational velocity is less than 1.15 \kms, which is consistent with our small $v\sin{i}$ estimate. 

Seven MCMC chains of 5$\times 10^5$ steps each is obtained to fit the observed data of $\alpha$ Cen B starting with random initial values within the following uniform priors:
\begin{eqnarray}
\alpha &=& [-\sqrt{S_{\mathrm{max}}},\sqrt{S_{\mathrm{max}}}] \\
\beta &=& [0,\sqrt{S_{\mathrm{max}}}] \nonumber\\
\gamma &=& [0,\sqrt{S_{\mathrm{max}}}] \nonumber\\
\mathrm{P}_{\mathrm{rot}} &=& [35,40]\nonumber\\
\mathrm{Jitter\,RV} &=& [0,10\times med(\sigma_{RV})],
\end{eqnarray}
where $S_{\mathrm{max}}$ is the maximum size allowed for the active region, fixed here at 20\%, and $med(\sigma_{RV})$ is the median of the RV error bars. Everything is similar to the fit done for HD189733, except that the RV, BIS SPAN and FWHM variations are fitted here and that the stellar radius, the limb darkening coefficients and the active region temperature are fixed to the values shown in Table \ref{tab:4}. In addition, only one jitter term is quadratically added to the RV, BIS SPAN and FWHM error bars when maximizing the log likelihood\footnote{For HARPS data, the error bars on the BIS SPAN and the FWHM are 2 and 2.35 times more than the ones for the RVs.}.
\begin{figure*}
\begin{center}
\includegraphics[width=16cm]{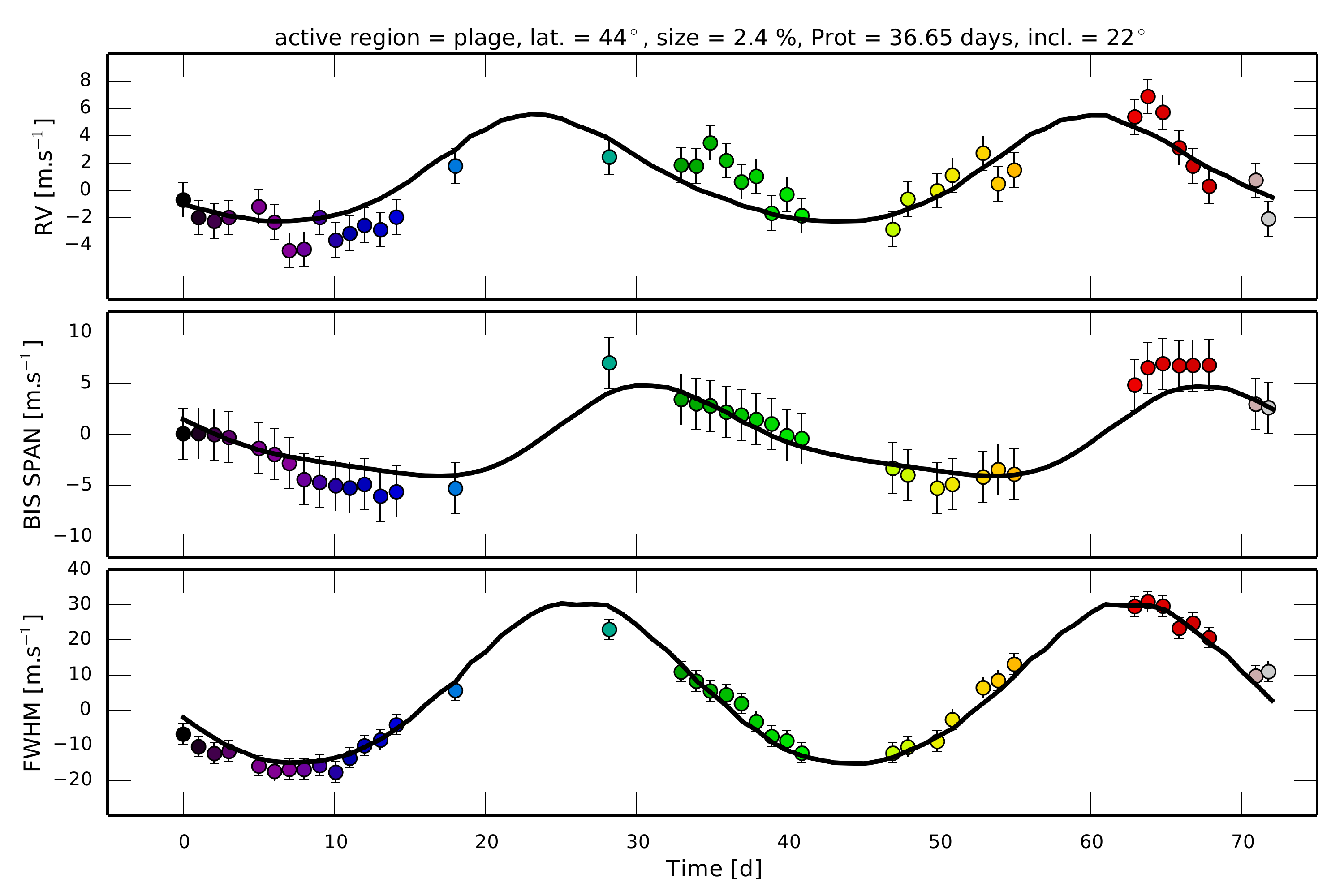}
\caption{RV, BIS SPAN and FWHM variations and best fit (black continuous line) of the activity-induced signal observed on $\alpha$ Cen B. The best-fitted solution corresponds to a plage of size 2.4\% that can be found at a latitude of 44 degrees. The star rotates in 36.65 days and is seen with an inclination of  22 degrees.}
\label{fig:3-3}
\end{center}
\end{figure*}
\begin{figure*}
\begin{center}
\includegraphics[width=16cm]{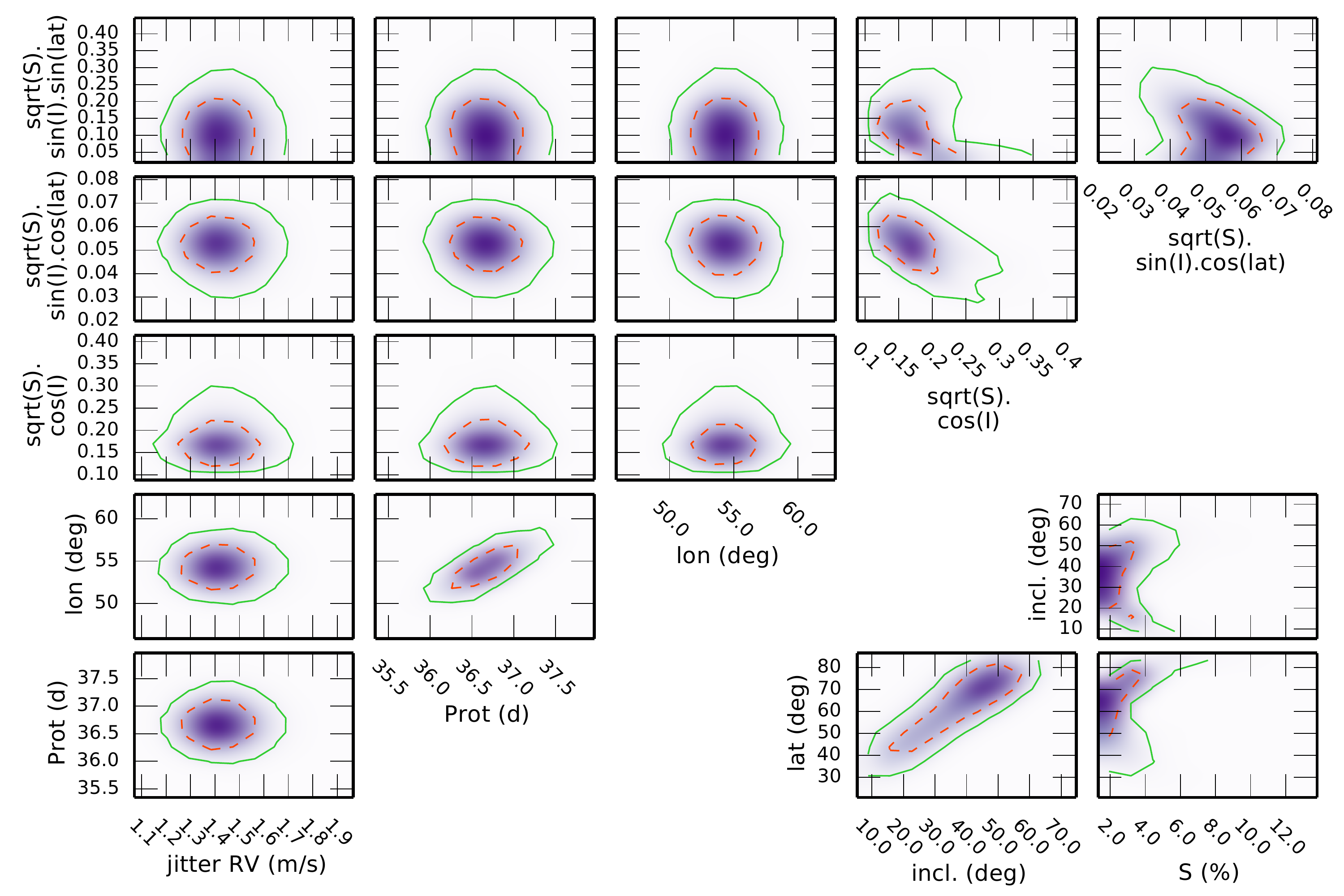}
\caption{Correlation between the different posterior distributions obtained from the MCMC fit to the data of $\alpha$ Cen B. The $1-\sigma$ and $2-\sigma$ contours are shown in dashed red and continuous green lines, respectively}
\label{fig:3-4}
\end{center}
\end{figure*}
\begin{figure*}
\begin{center}
\includegraphics[width=16cm]{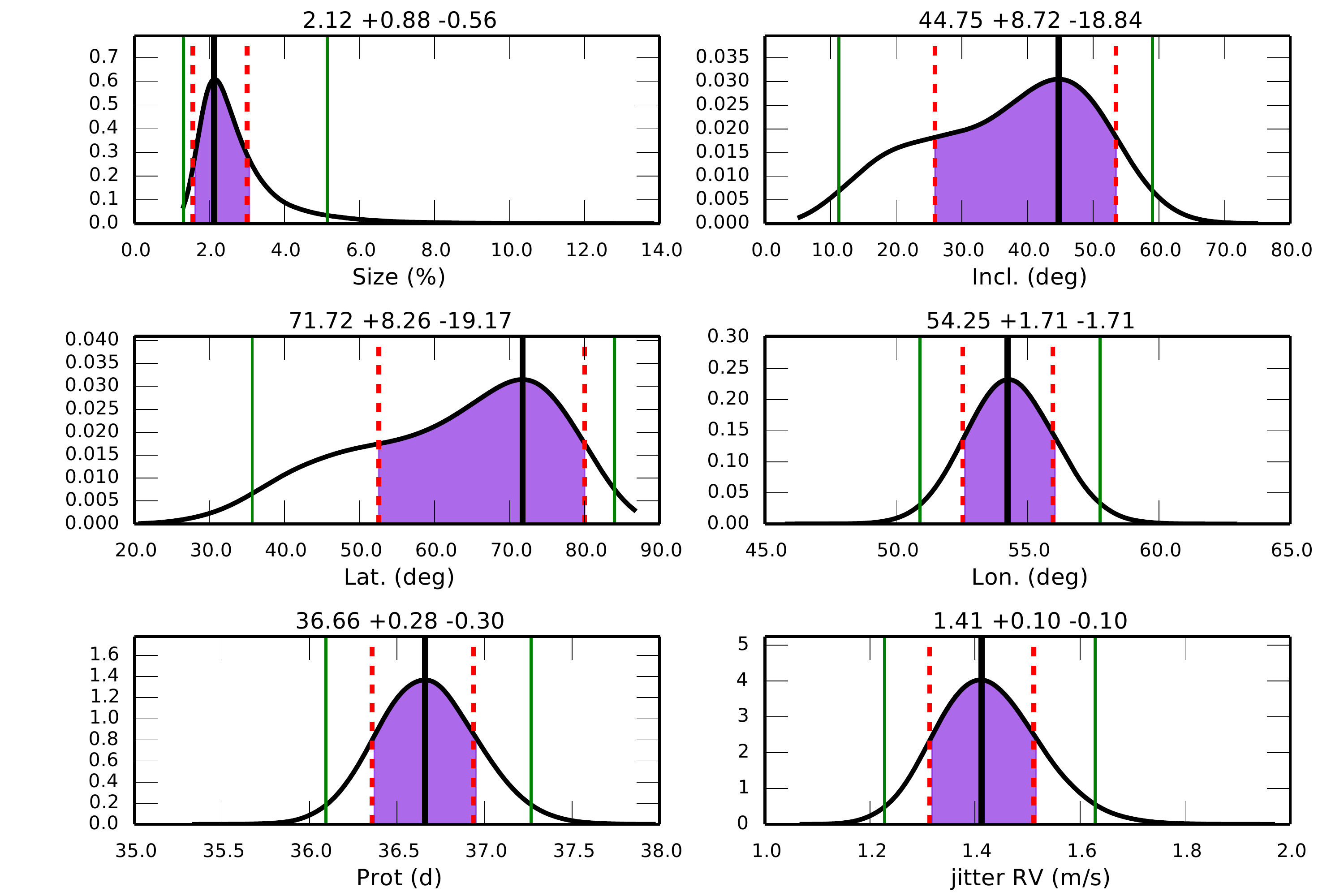}
\caption{Marginalized posterior distributions returned by our MCMC fit to the data of $\alpha$ Cen B. The mode (black thick line) of the distribution for each parameter, with its $1-\sigma$ uncertainty (purple shaded regions delimited by red dashed lines) and $2-\sigma$ uncertainty (delimited by green thin lines) can be found on each plot. The title of each plot gives the value for the mode of the distribution and its $1-\sigma$ uncertainty.}
\label{fig:3-5}
\end{center}
\end{figure*}
A Gelman-Rubin test \citep[][]{Gelman-2004} on the seven chains gives a potential scale reduction better than 1.0042 for all parameters, proving the convergence and the proper mixing of the chains. Following the positive result of the Gelman-Rubin test, we decided to run a long chain with 2$\times 10^6$ steps starting with initial values close to where the seven chains converged, and removed the first 2$\times 10^5$ steps to reject the burn-in-period. The correlation between the posterior distributions of the MCMC parameters are shown in Figure \ref{fig:3-4}. Figure \ref{fig:3-5} shows the marginalized posterior distributions for the size $S$, the latitude $\phi$ and the longitude of the active region, and the stellar rotational period and inclination $i$. The best-fitted model maximizing the log likelihood is shown by the black curve in Figure \ref{fig:3-3}, and corresponds to a configuration where the plage is at a latitude of $44$ degrees and has a size of $2.4$\%, on a star that have a stellar inclination of $22$ degrees. The reduced $\chi^2$ of this model is 1.00 compared to 11.17 for a flat model. Only considering the RVs, the reduced $\chi^2$ of the fit is 1.85 compared to 4.90 for a flat model, and the standard deviation of the RV residuals is 1.58\ms compared to 2.73\ms, respectively. Our best fitted model is therefore a better representation of the observed RV variations than a flat model. When comparing the stellar inclination and the active region latitude of the best fit (see Figure \ref{fig:3-3}) with the marginalized posterior distributions (see Figure \ref{fig:3-5}), we note that the best fitted solution is outside of the $1-\sigma$ uncertainty interval. This is not the case when looking at the correlation between the stellar inclination and the active region latitude in Figure \ref{fig:3-4}, where the best fitted solution is within $1-\sigma$. This discrepancy is induced by the marginalization of the posterior distribution on parameters that are correlated between each other.

$\alpha$ Cen B is one of the component of the $\alpha$ Cen binary system that have a nearly edge-on orbit relative to the line of sight with an angle of $79.20\pm0.04$ degrees \citep[][]{Pourbaix-2002}. Our measurement of the stellar inclination for $\alpha$ Cen B derived from the marginalized posterior $45^{+9}_{-19}$ excludes the spin-orbit of the star to be aligned with the binary orbital spin, with a difference greater than 20 degrees at $2-\sigma$. This misalignment is expected for wide separation binaries like the $\alpha$ Cen system \citep[][]{Jensen-2014,Hale-1994,Gillett-1988}. In addition, we also exclude the star to be pole on or equator on. Assuming a spin-orbit alignment for the close-in planet $\alpha$ Cen B\,b, our measurement of the stellar inclination implies that the planet is not transiting its host star. 

Simultaneous photometric measurements could constrain better the size of the active region, and therefore could improve the precision on each parameters. Unfortunately, such data with the required precision do not exist for $\alpha$ Cen B, and would be difficult to obtain because of the brightness of the star and therefore the lack of reference star to perform differential photometry\footnote{$\alpha$ Cen A could be used, however one reference star is often not enough for high photometric precision.}.

The rotational period of the star is estimated to $36.66^{+0.28}_{-0.30}$ days, which is one day faster than the value fitted in the discovery paper of $\alpha$ Cen B\,b \citep[$37.80\pm0.16$,][]{Dumusque-2012}. In this discovery paper, the activity was modeled by fitting sine waves at the rotational period of the star and its harmonics ($P_{\rm rot}/2$,$P_{\rm rot}/3$,$P_{\rm rot}/4$), which not always gives the correct rotational period estimate \citep[see Section 2.4 in][]{Dumusque-2011b}.


\section{Discussion and Conclusion}  \label{sect:6}

In this paper, we estimate the stellar inclination for a moderate and a slow rotator: HD189733 with $v\sin{i}\sim3$\kms and $\alpha$ Cen B with $v\sin{i}\le1.15$\kms, respectively.
This stellar inclination is derived with a new approach that uses the results of the SOAP 2.0 activity simulation to fit the photometric, RV, BIS SPAN and FWHM variations induced by stellar activity.
In the two examples shown in this study, in average 40 photometric and/or spectroscopic measurements covering two rotational period of the star are enough to recover the stellar inclination. This is much less that the number of measurements required to derive stellar inclination using asteroseismology, which is the only other technique to be able to measure inclinations for rotators slower than 2-2.5\kms, like it is the case for $\alpha$ Cen B.

In the case of HD189733, our estimate of the stellar inclination $i=84^{+6}_{-20}$ degrees can be used with a measurement of the the spin-orbit angle to obtain the obliquity of the star-planet system. We confirm that the obliquity is small, $\psi=4^{+18}_{-4}$ degrees, which was highly probable given the spin-orbit measurement that was very close to zero degrees \citep[e.g.][]{Triaud-2009}. In addition, we find that the active region responsible for the variation is a spot at a latitude of $67^{+12}_{-36}$ degrees, compatible with previous HST observations showing the occultation of spots by the transiting planet orbiting HD189733 \citep[][]{Pont-2007}.

For $\alpha$ Cen B, we find a stellar inclination of $45^{+9}_{-19}$ degrees, which excludes the rotational spin of $\alpha$ Cen B to be aligned with the orbital spin of the $\alpha$ Cen binary system. In addition, assuming that the close-in planet $\alpha$ Cen Bb is aligned with its host star, this estimate also exclude the transit of the planet. 

In the two examples shown in this paper, either we analyze good photometric measurements and poor spectroscopic observations, or precise spectroscopic data but without photometry. The combination of photometric measurements at the tens of ppm precision and spectroscopic data at the meter per second level would allow to better constrain the different stellar and active region parameters. The extended \emph{Kepler} mission (\emph{K2}), as well as \emph{CHEOPS} \citep[][]{Broeg-2013}, \emph{TESS} \citep{Ricker-2014} and \emph{PLATO} \citep{Rauer-2013} with the help of ground based spectrographs such as HARPS, HARPS-N, HIRES, ESPRESSO and G-CLEF will be able to deliver such data. Another alternative to simultaneous photometry would be to study the calcium activity index variation, which is obtained from spectroscopy and should be correlated with the photometric variation.

In our analysis, the stellar inclination can be obtained when one dominant active region is present on the stellar surface and if this active region evolves slowly in comparison with the stellar rotation period. The data from HD189733 and $\alpha$ Cen B studied in this paper show that this configuration of activity is possible, and therefore it is reasonable to think that other stars will show a similar behavior. As another example, when the Sun is at its maximum activity level, it is not uncommon to see one long-lived main active region on the stellar surface. 

An important point in our analysis is that the temperature of the active region is fixed to the solar value. Looking at stars different from the Sun, HD189733 and $\alpha$ Cen B are both K1V dwarfs, there is no reason why the active region temperature for these stars should be similar. In models trying to reproduce stellar activity, the active region temperature is always degenerated with the active region size at first order, because the signal of a big active region with a small contrast can be reproduced by a smaller active region with a higher contrast. However, the precise size and temperature of an active region are not important in our case, because the degeneracy between both does not significantly affect the estimation of the latitude of the active region, as well as the stellar inclination \citep[see discussion in e.g. ][]{Dumusque-2014b,Lanza-2009}.
We note however that in \citet{Pont-2013}, the temperature difference of a spot occulted during the transit of HD189733b is estimated to be $-750\pm250\,K$, compatible with our value of $-663\,K$ adopted here.
 
Only the stellar inclination for HD189733 can be compared to previous measurements. The next step will be to test if the stellar inclination found when fitting activity is compatible with other methods, like spectra fitting, CCF fitting, asteroseismology, or Zeeman-Doppler imaging (ZDI). This last method has the benefit of being able to infer a brightness map of the stellar surface for fast rotators, and therefore an idea on the position of the active region can be obtain \citep[e.g.][]{Donati-2013,Donati-2011,Skelly-2010}. An interesting test would be to see if there is a compatibility between the stellar inclination and active region latitude found by ZDI and by fitting stellar activity.

\acknowledgments
We thanks the anonymous referee for his valuable comments that improved the first version of the paper and made the results more significant. We are very grateful to the \emph{MOST} team for making available the observations analyzed in this work and for useful discussion. In addition, interesting discussion with R. Anderson, I. Boisse, D. Kipping, F. Pepe, S. Saar, N.C. Santos, D. S\'egransan and A. Triaud helped in improving the paper. We thank R. Anderson for a careful read of the paper and very helpful comments in return. X. Dumusque thanks the Swiss National Science Foundation for its financial support through an \emph{Early PostDoc Mobility} fellowship. 


\bibliographystyle{apj}
\bibliography{dumusque_bibliography}

\end{document}